# Exploring the Relationship between Urbanization and Ikization

Lü Ye [1], Yanguang Chen [2]

(1.College of Physical Education, Yangzhou University, Yangzhou, China, 225002; 2. Department of Geography, College of Urban and Environmental Sciences, Peking University, Beijing, 100871, China)

**Abstract**: The phenomenon of Iks was first found by anthropologists and biologists, but it is actually a problem of human geography. However, it has not yet drawn extensive attention of geographers. In this paper, a hypothesis of ikization is presented that sudden and violent change of geographical environments results in dismantling of traditional culture, which then result in collective depravity of a nationality. By quantitative analysis and mathematical modeling, the causality between urbanization and ikization is discussed, and the theory of replacement dynamics is employed to interpret the process of ikization. Urbanization is in essence a nonlinear process of population replacement. Urbanization may result in ikization because that the migration of population from rural regions to urban regions always give rise to abrupt changes of geographical environments and traditional culture. It is necessary to protect the geographical environment against disruption, and to inherit and develop traditional culture in order to avoid ikization of a nation. The approach to solving the problems caused by fast urbanization is to reconstruct geographical environment so that rural culture will be naturally replaced by urban culture.

## 1 Introduction

Urbanization is a complex dynamic process based on growth of urban population and migration of rural population. A process of urbanization includes two nonlinear geographical processes: one is gradual increase in the proportion between urban population and total population, and the other, is society changes of the people living in urbanized areas from a rural to an urban way of life (Knox, 2005; Zhou, 1995). The former is a quantitative change, and the latter is a qualitative change. The former indicates movement of rural population to the urban regions and concentration of urban people, while the latter is also known as urbanism, which is actually the characteristic way of interaction between urban inhabitants and geographical environment (Knox and Marston, 2009). For the urbanized people, there are two significant changes. One is the environmental change, that is, the rural environment is replaced by urban environment; and the other, cultural change, that is, the rural culture is substituted with urban culture.

As a whole, urbanization originates from industrialization. Without industrialization, there will be no urbanization in the modern sense (Knox and Marston, 2009). For a long time, the character of urbanization is regarded as positive. Seldom people think of it as a double-edged sword under certain conditions because of nonlinear dynamics. In fact, a theory study has shown that fast urbanization can cause periodic oscillation or even chaos of the proportion of people living in urban areas (Chen, 2009). The most terrible is that the bad way of urbanization can possibly give rise to ikization. The phenomenon of Iks was first found by anthropologists and biologists (Tainter, 2006; Thomas, 1974; Turnbull, 1972; Turnbull, 1978), but it is actually a problem of human geography (Chen, 2014a). However, it has not yet drawn extensive attention of geographers.

In this paper, a hypothesis of ikization is presented that sudden and violent change of geographical environments results in dismantling of traditional culture, which then result in collective depravity of a nationality. The replacement dynamics, a new theory of nonlinear systems, can be employed to explore the causality between ikization and fast urbanization. After all, urbanization is in essence a process of population replacement (Chen, 2014b). The rest of the article is arranged as follows. In section 2, a hypothesis of ikization will be put forward, and the causes and effects of ikization will be expounded. In section 3, the relationships between urbanization and ikization will be revealed. The causes of ikization lie in abrupt changes of geographical environments and traditional culture. In section 4, the theory of replacement dynamics will be employed to interpret the possible modes of ikization proceeding from urbanization. Moreover, the city of Beijing will be taken as an example to confirm the fast urbanization of China. The study will be concluded by a summarization. The novelty of this paper rests with three aspects. First, the relationships between ikization and fast urbanization of China are disclosed. Second, three conceptual models of ikization is presented. Third, the model of replacement dynamics is associated with the phenomena of ikization.

# 2 A hypothesis of ikization

## 2.1 The definition of ikization

First of all, the concept of ikization should be illuminated. Ikization is a phenomenon of human geography, but it was first discussed in anthropology and biology instead of geography. Chen (2014a) coined the word *ikization* in terms of the small tribe of Iks in Uganda. The people of Iks were formerly nomadic hunters and gatherers in the mountain valleys of northern Uganda, a country of east-central Africa. However, because the government decided to construct a national park, that is, Kidepo Valley National Park, they were compelled by law to give up hunting in the valleys and become farmers on poor hillside soil (Turnbull, 1972). Since then, the tribe of Iks became a mean society (Thomas, 1974): "*These people seem to be living together, clustered in small, dense villages, but they are really solitary, unrelated individuals with no evident use for each other. They talk, but only to make illtempered demands and cold refusals. They share nothing. They never sing. They turn the children out to forage as soon as they can walk, and desert the elders to starve whenever they can, and the foraging children snatch food from the mouths of the helpless elders….. They breed without love or even casual regard. They defecate on each other's doorsteps. They watch their neighbors for signs of misfortune, and only then do they laugh.*" A British-American anthropologist, Colin Turnbull (1972), found this tribe, and wrote a controversial book entitled "The Mountain People". This book interested scientists (Thomas, 1974), but also attracted criticism from author's peers (Heine, 1985). Some scholars regarded that "this book cannot be discussed in any proper sociological terms, for we are provided with only snatches of data. " (Beidelman,1973; Heine, 1985) Despite this objection, the book made the Ik people famous and become "*literary symbols for the ultimate fate of disheartened, heartless mankind at large*" (Thomas, 1974).

The misfortune of Iks suggests that a tribe's ikization arises from radical change of geographical environments and break of traditional culture. Sudden and violent changes of geographical environments may induce discontinuity of inherent culture. Cultural rupture will bring about loss of people's sense of place and result in collective depravity of a community or even a nationality. Actually, as indicated by the biologist, Thomas (1974), "*the Iks have transformed themselves into an irreversibly disagreeable collection of unattached, brutish creatures, totally selfish and loveless,* *in response to the dismantling of their traditional culture.*" There are two concepts that can be employed to describe the process of ikization: changes of geographical environment and cultural breaking (Chen, 2014a; Tainter, 2006; Turnbull, 1978). Both the environment variation and culture variation are involved with two aspects: one is absolute change, namely, devastation, and the other is relative change, that is, migration.

Now, a conceptual model can be built for our understanding the causality of ikization. If the degree of ikization is treated as a function (output variable, dependent variable, response variable), environmental variance and cultural break can be regarded as two arguments (input variables, independent variables, explanatory variables). Thus the causality can be expressed as a linear regression equation as below:

$$\text{Ikization} = a + b_1 \times (\text{environmental change}) + b_2 \times (\text{cultural break}), \tag{1}$$

where $a$ refers to an intercept (constant), and $b_1$ and $b_2$ to regression coefficients reflecting the impact strength. Alternatively, the relationship between cause and effect of ikization can also be described by a production function, which is simple nonlinear regression equation, as follows

$$\text{Ikization} = a \times (\text{environmental change})^{b_1} \times (\text{cultural break})^{b_2}, \tag{2}$$

where $a$ denotes a constant, and $b_1$ and $b_2$ are two elasticity coefficients (Luo and Chen, 2014).

No human culture is independent of its geographical environment. Human being depends on earth surface, and culture results from the interaction and ecological relationships between human being and physical environment. In this sense, equation (2) is more preferable than equation (1), but equation (1) is simpler and thus easier to understand than equation (2). In fact, we can find the third function to model the causality about geographical processes and ikization, which will be discussed in the fourth section.

## 2.2 Political and economic movements resulting in ikization

If we read the modern history of China, we can find that many events have brought about cultural catastrophes and environmental and ecological disasters. In some cases, a good motivation may lead to bad results. This implies the nonlinearity and thus complexity of social and economic systems. Today, the symptom of ikization has appeared among parts of Chinese people after a long latent period, which suggests the effect of time lag between the causes and effects of a geographical process. Since the May Fourth Movement (1919), which was correlated with the New Culture Movement (1915-1923), Chinese traditional culture has been questioned, reconsidered, or even denied. This movement was regarded as an anti-imperialist and anti-feudal political and cultural movement and thought to be a historical progress of China, but its aftereffects are very complicated. During the Great Leap Forward and the People's Commune Movement (1958-1960), which represents the high point of ignorant mass folly, Chinese geographical circumstances and conditions as well as ecological systems suffered serious destruction (Cannon and Jenkins, 1990). Especially, the Great Proletarian Cultural Revolution (1966-1976) gave rise to both environmental and cultural devastation. Since the introduction of the policies of reform and opening-up at the end of 1978 and with the gradual establishment of a socialist market economic system from 1992, namely, after Deng's South Tour Speeches, China's national economy and cities developed rapidly (Yang, 1998). However, this development once inflicted heavy losses of geographical environments due to absence of a rule of law (Table 1). In recent years, the Chinese government has finally realized the importance of protecting the geographical environment. A series of new policies and measures have been introduced, which have significantly improved China's natural environment and ecological conditions.

**Table 1 Important historical events associated with great political and economic movements resulting in ikization of Chinese**

| Event | Time | Related concept | Consequence |
|---|---|---|---|
| The May 4th Movement | 1919 | New Culture Movement | Reconsider Chinese traditional culture |
| Great Leap Forward | 1958-1960 | The People's Commune Movement | Environmental and ecological destruction |
| Great Cultural Revolution | 1966-1976 | Ten Chaotic Years | Environmental and cultural destruction |
| South Tour Speeches | 1992 | Reform and Opening-up | Geographical environmental destruction |
| Fast Urbanization | Recent years | House Demolition and City-Making Movement | Geographical environmental destruction |

What is worth mentioning is that political and economic corruption represents a significant factor contributing to national ikization. In ancient China, corruption is an important ruling means used by almost all monarchs. If an emperor tried to plunder the national wealth through a state policy, he would be regarded as a bandit and would result in loss of morale of nations. A foxy emperor never easily robbed the people of their wealth. Instead, he would rather utilize the corrupt officials to plunder the wealth of the people. If a corrupt official knew when to stop and just leave, he will not be punished in accordance with the law. However, if he was excessively greedy of gain and have suffered the bitter hatred of the people, he would be chastised in the name of the law. His house would be searched and his property, of course, the mammon of unrighteousness, would be confiscated and finally obtained by the emperor. The emperor would kill two birds with one stone by punishing the corrupt officials and family members seriously. First, he acquired the wealth from his subjects by an indirect means. Second, he was regarded as a wise monarch and won the support of the people. For the emperor, the corrupt officials had two benefits. First, they were easily controlled because their handles were caught by the emperor. Second, they amassed huge wealth by plundering the national people for his monarch. What with the drawbacks of political institutions and what with the weakness of human nature, the corrupt officials emerged in an endless stream during the long process of Chinese history.

In modern society, the source of political corruption is the very source of unsustainable development. The ancient emperors didn't care about official corruption, which, as indicated above, was an approach for the emperors to make a fortune. After all, the officials could not emigrate abroad, so they would not transfer the ill-gotten gains to the overseas. What is more, due to the technical level, the extent of damage to the geographical environments caused by the official corruption was small. However, today, the situation is different from that in ancient times. In order to acquire immense amount of treasure through exploitation of natural resources, the corrupt officials always undermine environments crazily by colluding with merchants. Malfeasant official behaviors rely heavily on altisonant approaches such as reservoir resettlement and fast urbanization. Because of the advanced technology, it is easy for officers and businessmen to destroy the natural environments. In particular, they should not be responsible for the land as both the corrupt officials and illegal dealers can immigrate overseas. In a sense, modern corruption can cause great harm to the geographical environments and the relations between man and land, which lead to ikization of people, which in turn aggravate official corruption. Therewith a vicious circle comes into being, wrecking the country and bringing ruin to the people (Figure 1).

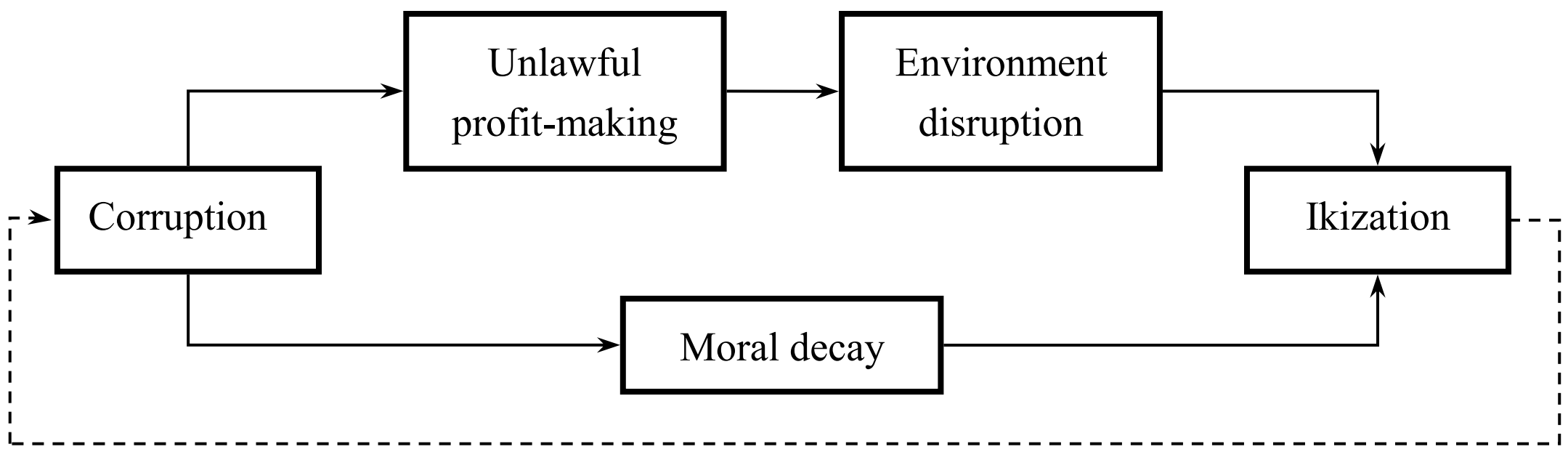

**Figure 1 The causal sequence from official corruption to ikization and in turn to corruption**

# 3 Urbanization and ikization

## 3.1 The dynamics of urbanization

Urbanization may be one of significant factors of ikization in the history of mankind. Urbanization implies the process of migration of rural populations into cities and towns, which results in an increasing proportion of the regional population residing in urban settlements. A city and a network of cities are complex spatial systems (Albeverio *et al*, 2008; Batty, 2005; Batty, 2013; Bertuglia *et al*, 1998; Wilson, 2000), and urbanization is a self-organizing process of urban evolution (Allen, 1997; Batty and Xie, 1999; Haken and Portugali, 1995; Portugali, 2000). In a sense, complex spatial patterns and self-organized processes represent the two different sides of the same coin of city development. The dynamics of urbanization can be understood by two processes. First, urbanization indicates a kind of *phase transition* from a rural to an urban settlement system (Anderson *et al*, 2002; Sanders *et al*, 1997). Phase transition is originally a physical term, which is often associated with the concept of self-organized criticality (SOC) (Bak, 1996; Batty and Xie, 1999). In physics, a phase transition denotes the transformation of a thermodynamic system from one state (e.g., liquid) of matter to another one (e.g., solid or gaseous) by heat transfer (Stanley, 1971; Yeomans, 1992). Today, the term can be employed to describe an evolution of a geographical system from rural state into urban state by urbanization. Second, urbanization suggests a complex *replacement dynamics*: rural settlements are replaced by urban settlements, and rural population is replaced by urban population (Chen, 2012; Chen, 2014b; Rao *et al*, 1989). In essence, phase transition and replacement dynamics also represent the two different sides of the same coin of urbanization.

In geography, urbanization involves both quality and quantity. A common knowledge is that the process of urbanization involves four aspects: urban system, urban form, urban ecology, and urbanism (Knox, 2005; Knox and Marston, 2009). Both urban system and urban form can be well described from the prospective of natural cities (Jiang and Miao, 2015; Jiang *et al*, 2015). However, it is difficult to study urban ecology and urbanism through modern technology. Among the four aspects, urbanism reflects the quality of urbanization and is more correlated with ikization. The term "urbanism" suggests that the culture or way of rural life of the urbanized people is substituted with the urban culture or way of urban life. The ikization can be associated with urbanization because that the transfer of population from rural regions to urban regions may give rise to abrupt changes of geographical environments and traditional culture.

## 3.2 Urbanization and ikization

For the city dwellers, the phase transition of urbanization includes two replacement processes during rural-urban population migration. First, the rural geographical environment is replaced by urban geographical environment. Second, the rural culture of life way is replaced by urban culture or life way. Thus the original sense of places will be lost. It will take a long time for the migratory dwellers to form new sense of places. This may result in ikization (Figure 2). In fact, Thomas (1974) once pointed out: "*Cities have all the Ik characteristics. They defecate on doorsteps, in rivers and lakes, their own or anyone else's. They leave rubbish. They detest all neighboring cities, give nothing away. They even build institutions for deserting elders out of sight.*" This suggests that urban inhabitants look like Iks because of environmental and cultural changes.

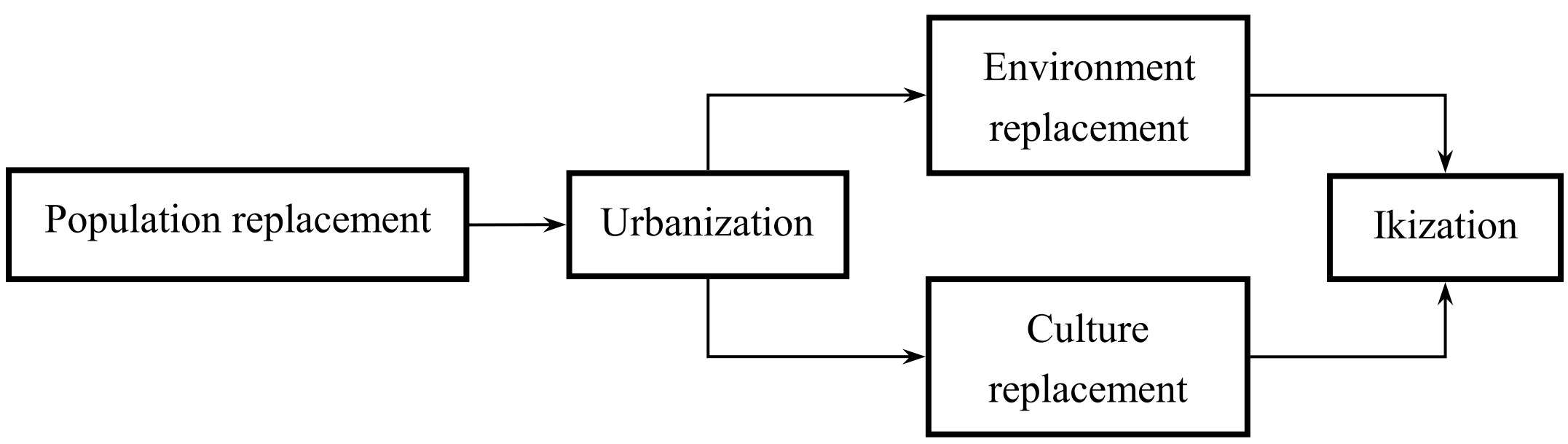

**Figure 2 The replacement dynamics of urbanization which may result in ikization**

It is rapid urbanization instead of natural urbanization that results in ikization of a community or even a nationality. In China, ikization may proceed from fast urbanization because of large-scale demolition of traditional urban communities and rural settlements. A new word, *chaiqian*, has emerged as the times require. Chaiqian means remove and relocation, that is, dismantlement of houses and movement of local residents. In many regions, government officials and business owners acted in collusion with each other to be engaged in rent-seeking of land. Many traditional settlements are pulled down, and local residents were compelled to move to the appointed places and were relocated. Many people became landless peasants coming between urban state and rural state and form a kind of marginal people living at the edge of modern society. Their situations were similar to Iks. Overnight, their life way was thoroughly remodeled. In short, their living environments were suddenly changed, and their inherent culture was broken down. They felt that they were rapidly transferred from one time and space to another time and space.

# 4 Theoretical analyses

## 4.1 Logistic replacement

The process of urbanization is associated with spatial replacement dynamics. Replacements are ubiquitous phenomena in both nature and society, which always take on a sigmoid curve (Batten, 1982; Chen, 2012; Fisher and Pry, 1971; Hermann and Montroll, 1972; Karmeshu *et al*, 1985; Montroll, 1978). In the simplest case, the replacement dynamics can be modeled with a logistic function. Among various replacement processes, urbanization and urban growth are two kinds of typical geographical substitution. Urbanization is a process of urban-rural population replacement, which has been studying for many years (Chen, 2014b; Rao *et al*, 1989). A new finding is that urban growth represents a complex dynamics of spatial replacement, which can be modeled with Boltzmann's equation and logistic function (Chen, 2012). Urban replacement is one of the nonlinear processes of geographical replacement. More important geo-replacement is that natural environment is replaced by human systems by degrees. In geography, the very basic and significant topic is the nonlinear relation between man (human societies) and land (natural environment) (Martin, 2005). The man-land relation can be regarded as the most important ecological relation in this world. Because of the interaction between human being and the surface of Earth, the primary productivity or even the secondary productivity are gradually consumed by humankind. So far, man has used up more than 40% of the primary productivity. In other words, human being has transformed the first nature into the second nature and the third nature (Chen, 2014b; Swaffield, 2002). To characterize this replacement process, a new measurement, the ratio of the primary productivity to the total productivity in a region, $p(t)$, can be defined by

$$p(t) = \frac{x(t)}{x(t) + y(t)} \times 100\% , \tag{3}$$

where $y$ represents the primary productivity, and $x$ is the other productivity. Accordingly, the ratio of the other productivity to the primary productivity (POR) can be defined as $o(t)=x(t)/y(t)$ (Chen, 2014b). Based on equation (3), a logistic model of man-land replacement can be built, and its mathematical expression is as below:

$$p(t) = \frac{1}{1 + (1/p_0 - 1)e^{-kt}} , \tag{4}$$

in which $p_0$ is the initial value of the productivity ratio $p(t)$, i.e., the ratio of productivity of time $t$=0, $k$ is the original rate of growth.

Because of population explosion, natural space has been rapidly replaced by human space all over the world. In fact, human race depletes geographical environment so fast by predatory exploitation of natural resources that the man-land relation becomes unstable. If we look at the earth from space, we will be surprised at its change. In particular, China seen from the satellite is like a piece of withered and yellow leaf rather than a dark green leaf (Yu, 2006). We cannot find a place in our country as Land of Peach Blossoms, i.e., land of idyllic beauty. If things continue this way, periodic oscillations or even chaos may arise some day in the future (Chen, 2014b). If the ratio of the other productivity to the total productivity is substituted by the level of urbanization, $L(t)$, we will have an logistic model of urban-rural replacement, that is

$$L(t) = \frac{1}{1 + (1/L_0 - 1)e^{-kt}} , \tag{5}$$

where $L_0$ is the initial value of the urbanization level $L(t)$ at the time $t$=0, $k$ is the inherent rate of growth of the ratio of urban population to total population.

### 4.2 Step replacement and ikization

Logistic replacement is a natural replacement, which cannot result to ikization. Fast urbanization curve should be described with quadratic logistic function, and can be termed quadratic logistic replacement. Both logistic replacement and quadratic logistic function belong to sigmoid replacement. However, another type of geographical replacement, step replacement, will give rise to ikization. The so-called ikization is in fact a process of human replacement: noble people will be substituted by mean people (Figure 2). The step replacement is a dynamic process that can be characterized by the unit-step function

$$p(t)=\begin{cases}0, & t<t_0 \\ 1, & t\geq t_0\end{cases}, \tag{6}$$

where $t_0$ denotes a critical time. The unit-step function can be treated as the extreme special case of the logistic function, or the logistic function can be thought of as a smooth approximation to the unit-step function. Actually, in equation (4), if $t$=0, we will have $p(t)=p_0=0$, while if $t\to\infty$, then it follows $p(t)$=1. Therefore, this replacement can also be termed 0-1 replacement. The unit-step function can be employed to model the processes of substitution such as reservoir resettlement and large-scale dismantlement of houses and movement of local residents. For urbanization, three types of replacement dynamics can be tabulated and illustrated as below (Table 2, Figure 3). During urbanization, the original sense of places of the urbanized people gets lost, and it will take a long time for the migratory dwellers to form a new sense of place. The loss of place sense may be one of influencing factors of ikization.

**Table 2 Three types of urbanization replacement and the corresponding mathematical models**

| Urbanization | Replacement | Model | Feature |
|---|---|---|---|
| Natural urbanization | Natural replacement | Logistic function | Gradual change |
| Fast urbanization | Fast replacement | Quadratic logistic function | Rapid change |
| Dismantlement/movement | Step replacement | Unit-step function | Sudden change |

In theory, ikization can be measured with 0 and 1. If ikization appears, it is 1, or else it is 0. Thus the ikization process can be depicted with a unit-step function, equation (6). Based on the step replacement, the conceptual model of ikization can be substituted with a logistic regression model in the following form

$$\text{Ikization} = \frac{1}{1+ae^{-[b_1\times(\text{environmental change})+b_2\times(\text{cultural break})]}}, \tag{7}$$

where $a$, $b_1$, and $b_2$ are parameters. In this instance, the degree of ikization will be measured with a dummy variable rather than a metric variable. A dummy variable is also known as a binary variable, Boolean indicator, categorical variable, design variable, indicator variable, or qualitative variable, which is always represented by 0 and 1 (Diebold, 2004; Kleinbaum *et al*, 1998). Now, we have three possible mathematical models to describe the causality of ikization (Table 3).

**Table 3 Three possible conceptual models of ikization indicative of causes and effect**

| **Relation** | **Ikization measure** | **Function** | **Model** |
|---|---|---|---|
| Linear relation | Metric variable | Linear function | $y = a + b_1x_1 + b_2x_2$ |
| Nonlinear relation | Metric variable | Production function | $y = ax_1^{b_1}x_2^{b_2}$ |
| | Dummy variable | Logistic function | $y = \frac{1}{1+ae^{-(b_1x_1+b_2x_2)}}$ |

**Note**: In the table, $y$ denotes the degree of ikization, $x_1$ refers to environmental change, and $x_2$ to cultural break. As indicated above, $a$, $b_1$, and $b_2$ are three parameters.

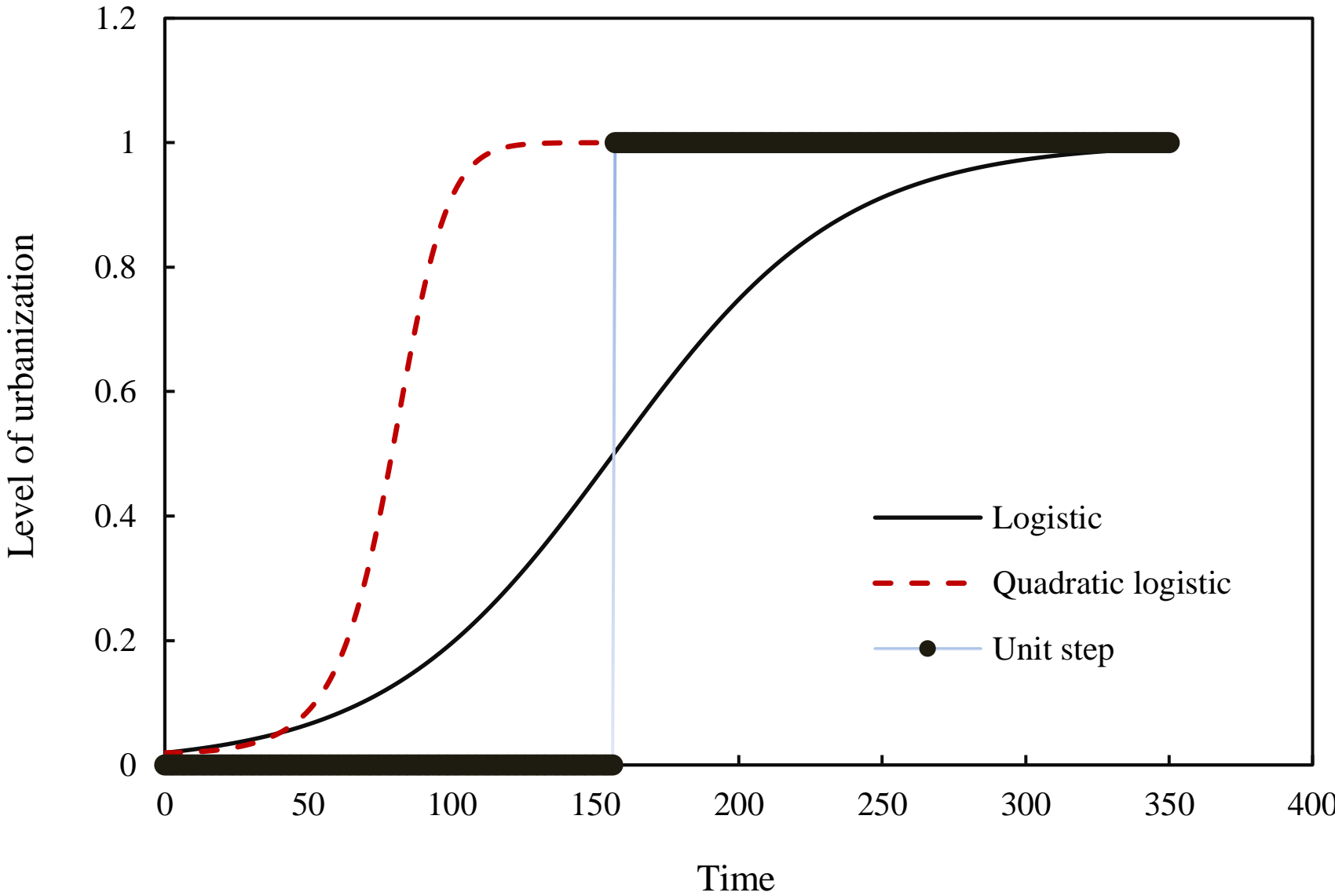

**Figure 3 Three types of urbanization curves indicating three types of replacement dynamics**

### 4.3 The case of Beijing

Because of the destruction of geographical environment and the rupture of traditional culture, the symptom of ikization seems to have appeared in partial Chinese people. Various phenomena of Iks such as corruption, indifference, schadenfreude, no sense of responsibility, and making counterfeit goods have been discussed in detail (Chen, 2014a). On the other hand, the problems of fast urbanization of China have been studied by many scholars from varied perspectives. However, the mathematical modeling and quantitative analyses are seldom reported on rapid urbanization in literature. In China, the typical signs of fast urbanization are urban sprawl, bubble economy of real estate, and large-scale demolition of traditional settlements. These signs are correlated with one another. As indicated above, one of aspects of urbanization is urban form, which takes on sprawl during the stage of rapid urbanization. Beijing, the capital of China, can be regarded as a microcosm of China, which in turn can be treated as a macrocosm of Beijing. The models of a city's growth is always consistent with the model of urbanization of its country. We might as well take Beijing as an example to illustrate the recent state of China's fast urbanization.

The basic tools of replacement dynamics including squashing functions, sigmoid curves, and allometric equation can be utilized to analyze the rapid growth of Beijing city. The measurements comprise the traditional measures, urban area and population size, and the new measurement, scaling exponent. In theory, urban form is free of characteristic scale and cannot be described with common measures such as area and size. Urban area can be subjectively defined rather than objectively measured. At present, three approaches can be employed to designate metropolitan areas or demarcate urban agglomerations. The first is the city clustering algorithm (CCA) presented by Rozenfeld *et al* (2008, 2011), the second is the fractal-based method presented by Tannier *et al* (2011), and the third is the variant of CCA based on street nodes/blocks developed by Jiang and Jia (2011). In this work, the metropolitan area of Beijing is determined by using CCA. By means of eight years of remote sensing data of urbanized area from 1984 to 2009 and the six times of census data of urban population from 1953 to 2010, we can examine the urban sprawl of Beijing in the past thirty years.

The squashing functions include the conventional logistic function, fractional logistic function, and quadratic logistic function. The urbanized area of a city can be modeled with a fractional logistic function (Chen, 2014c). Based on the dataset consisting of 8 data points from 1984 to 2009 (Table

4), a model of Beijing's area growth can be built as below

$$\hat{A}(t)=\frac{3881716000}{1+9.0686e^{-(0.0829t)^{4/3}}}, \tag{8}$$

where $t$ denotes time, and $A(t)$ refers to the urbanized area of time $t$. The goodness of fit is about $R^2$=0.9964 (Figure 4). In terms of this model, the inherent rate of growth of urban area is approximately 0.0829.

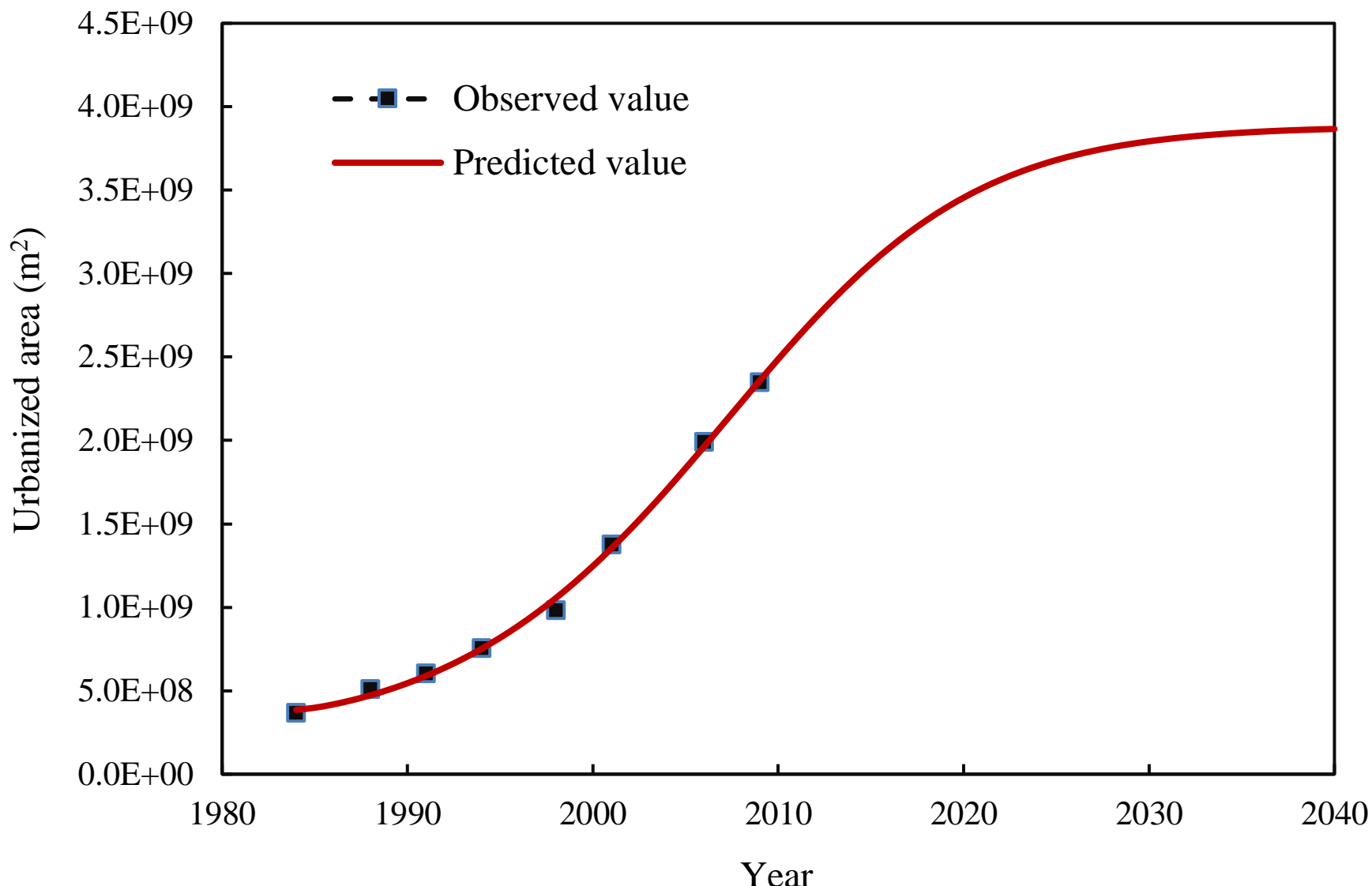

**Figure 4 The fractional logistic curve of urbanized area growth of Beijing city (1984-2040)**

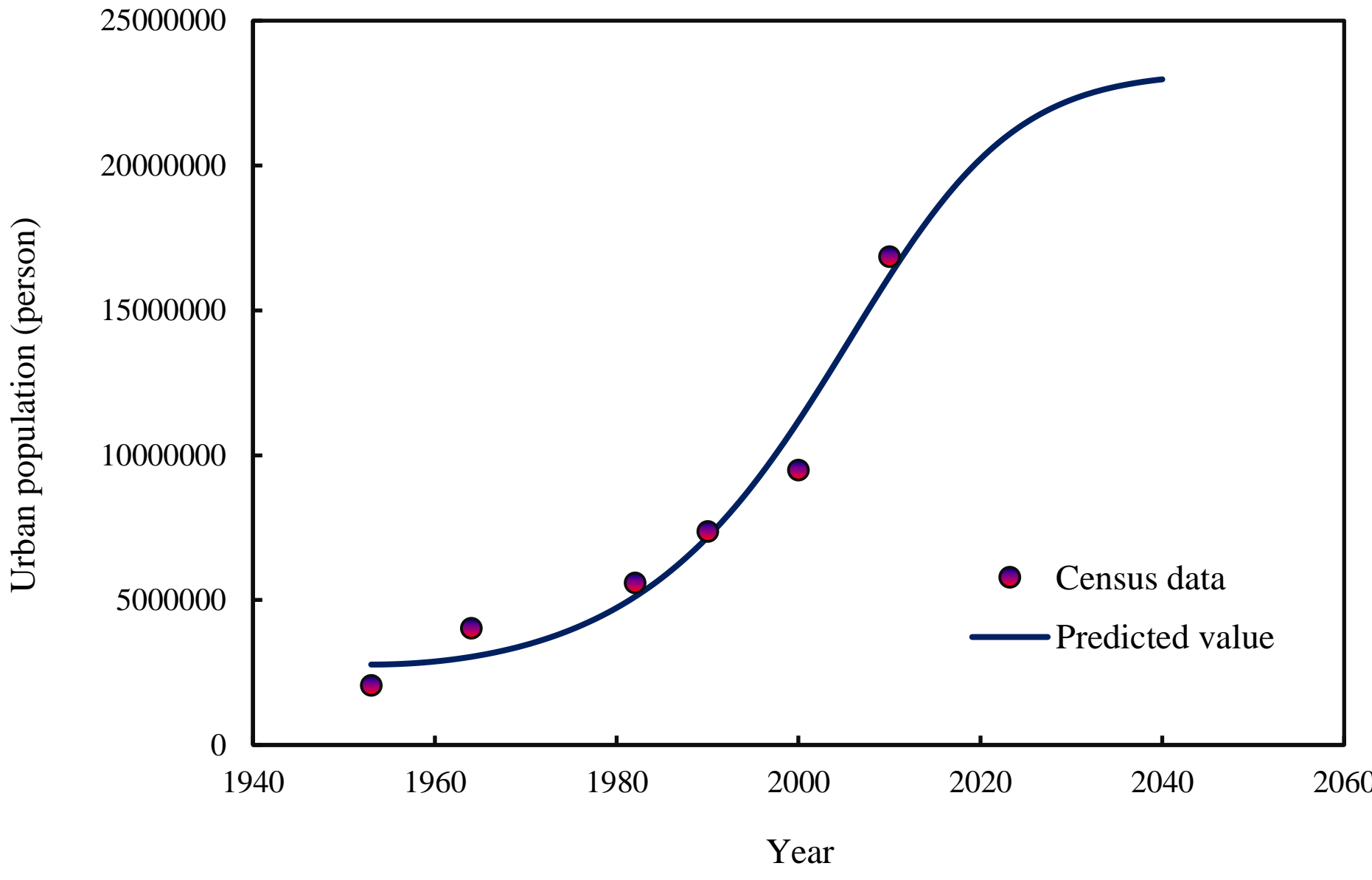

**Figure 5 The quadratic logistic curve of urban population growth of Beijing city (1984-2040)**

Where Beijing is concerned, there are both similarities and differences between urban area and urban population. The population growth of Beijing can be modeled with a quadratic logistic function (Chen, 2014b). Based on the dataset comprising six times of census data, a model can be made as follows

$$P(t)=\frac{23206725}{1+7.3611e^{-(0.0295t)^2}}, \tag{9}$$

where $P(t)$ denotes the population size of time $t$. The goodness of fit is about $R^2$ = 0.9459 (Figure 5). In light of this model, the inherent rate of growth of urban population is around 0.0295. The relative growth rate of urbanized area (0.0829) is significantly greater than that of urban population size (0.0295). Under condition of normal urbanization, the relative growth rate of urban area is supposed to be less than that of urban population.

By means of exponential interpolation, we can complement the population dataset of Beijing. Then an allometric analysis can be made for the scaling relation between urbanized area and urban population. By the least squares calculations, an allometric growth model is obtained as below

$$A(t)=0.0000003109P(t)^{2.2176}. \tag{10}$$

The goodness of fit is about $R^2$ = 0.9872 (Figure 6). According to this model, the allometric scaling exponent is approximately 2.22, which is significantly greater than 1. The allometric scaling exponent is the ratio of the relative growth rate of urban area to that of urban population:

$$\sigma=\frac{\mathrm{d}A(t)/A(t)}{\mathrm{d}P(t)/P(t)}\approx 2.2176, \tag{11}$$

where $\sigma$ denotes the allometric exponent. Generally speaking, the allometric scaling exponent comes between 2/3 and 1 (Chen, 2014c). Once again, the result shows that the relative growth rate of urbanized area is greater than that of urban population size. This suggests the waste phenomenon in the land use of Beijing. Obviously, the conclusion drawn from equations (10) and (11) lends further support to the conclusion drawn from equations (8) and (9).

In theory, a city has no characteristic scale and urban form and growth should be described with fractal dimension. Fractal dimension is a kind of scaling exponent indicating the extent of space filling and the degree of spatial complexity. The fractal dimension can be evaluated with the box-counting method, and the fractal dimension growth can be modeled with sigmoid functions (Chen, 2012). Based on the eight years of observational data (Table 4), the fractal dimension growth of Beijing can be modeled by a quadratic logistic function (Chen, 2018). The quadratic logistic growth comes between exponential growth and the conventional logistic growth. The quadratic logistic model suggests that the relative growth rate of urban sprawl is much higher than the relative growth rate under normal circumstances. As indicated above, a city's growth is consistent with urbanization of a nation. The model of fractal dimension growth of Beijing's urban form lends further support to the suggestion of fast urbanization of China.

**Table 4 The urban area, fractal dimension, and the related measurements of Beijing city (1984-2009)**

| Year | Time square ($t^2$) | Urbanized area ($A$) | Fractal dimension ($D$) | Filled-unfilled ratio (FUR) | Level of space filling (LSF) |
|---|---|---|---|---|---|
| 1984 | 0 | 365058785.8 | 1.4280 | 0.9846 | 0.4961 |
| 1988 | 16 | 508757623.1 | 1.4465 | 1.0729 | 0.5176 |
| 1991 | 49 | 601060517.8 | 1.4931 | 1.3342 | 0.5716 |
| 1994 | 100 | 752786609.9 | 1.5074 | 1.4279 | 0.5881 |
| 1998 | 196 | 980603624.8 | 1.5967 | 2.2424 | 0.6916 |
| 2001 | 289 | 1375795387.0 | 1.6582 | 3.2170 | 0.7629 |
| 2006 | 484 | 1990590617.0 | 1.7607 | 7.4584 | 0.8818 |
| 2009 | 625 | 2347738581.0 | 1.7888 | 10.6706 | 0.9143 |

**Note**: The measurement filled-unfilled ratio (FUR) and level of space filling (LSF) are defined by Chen (2012). See also Chen (2018).

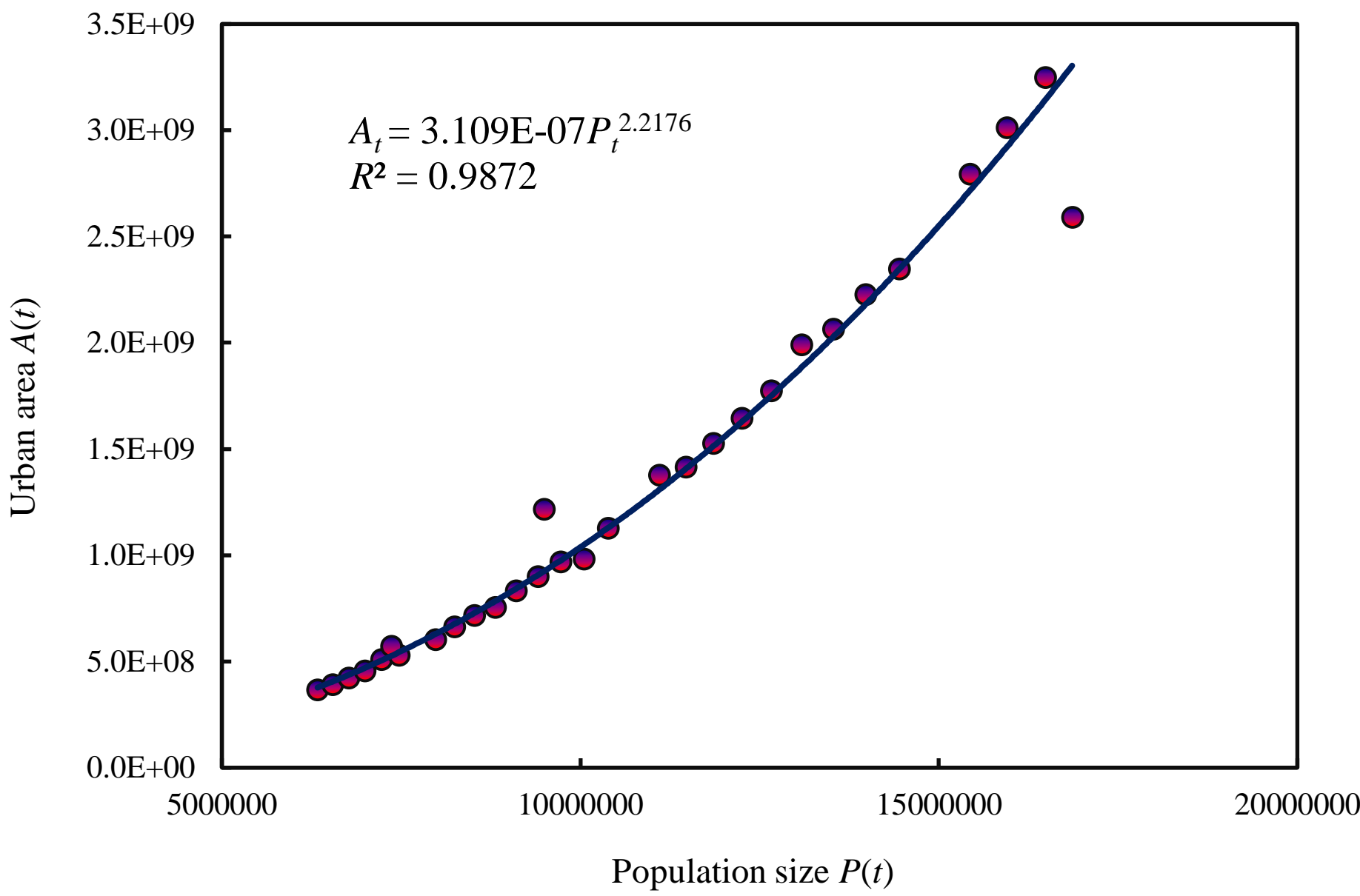

**Figure 6 The allometric scaling relation between urban population and urbanized area of Beijing**

**city (1984-2013)**

Now, based on mathematical models and model parameters, two criteria associated with replacement dynamics can be used to verify fast urbanization. One is the scaling exponent of allometric growth model, and the other is the latent scaling exponents of sigmoid models. The natural urbanization curve can be described with the conventional logistic function. If the urbanization curve exhibits a quadratic logistic function, the latent scaling exponent is 2, and thus it indicates fast urbanization. If the urbanization curve takes on a fractional logistic function, and the latent scaling exponent is significantly greater than 1, it suggests fast urbanization. In particular, if the allometric scaling exponent is greater than 1, which suggests the relative growth rate of urban area is significantly greater than that of urban population, it implies fast and immoderate urbanization. Beijing is a megacity in the world. If the characters of fast urbanization appear in the middle-sized and small cities, the thing can be regarded as normal. However, a megacity such as Beijing shows varied signs of fast urbanization, it is abnormal. This suggests a sort of urbanization bubble or bubble economy behind urbanization, which in turn suggests environment disruption and cultural rupture.

# 5 Conclusions

In this paper, the concepts of urbanization, ikization, and replacement dynamics are associated with one another. The main points of this study are as follows. First, ikization is connected with nonlinear man-land relationships. The causes of ikization rest with sudden change of geographical environment and break of traditional culture, and the consequence of ikization is national degeneration. Environmental changes lead to loss of sense of place, which give rise to collective depravity of a nationality. Second, urbanization results in rapid change of geographical environment and original culture. Because of migration from rural places to urban places, people's living environment will be relatively changed, and meanwhile the rural culture will be suddenly substituted by urban culture. Where the urbanized population is concern, the original sense of place will disappear overnight. Third, the replacement dynamics can be employed to interpret the ikization resulting from urbanization. Replacement processes can be divided into three groups: natural replacement, rapid replacement, and step replacement. Natural replacement model can be used to describe European and American urbanization, rapid replacement model can be used to reflect China's urbanization, and step replacement can be used to depict large-scale dismantlement of houses and movement of local residents. It is the step replacement rather than the natural replacement that results in ikization of people in a region or subregions.

There is always a time-lag relation between causes and effects in the evolution of complex systems. The past causes lead to the present and future effects, and the present causes lead to the future effects. Where there is time lag, there is nonlinearity; and where this is nonlinearity, there is complexity. The madness of men not only brings about the revenge of nature on human beings but also results in the fall of men themselves. The large-scale demolition of traditional urban communities and rural settlements in the process of fast urbanization suggests more adverse outcomes related with ikization in the future. It is necessary to protect the geographical environment against disruption, and to inherit and develop traditional culture in order to avoid ikization of a nationality. One of the important approaches to solving the problems caused by fast urbanization is to safeguard and reconstruct geographical environment so that rural situations and culture will be naturally replaced by urban situations and culture. Fortunately, Chinese government has realized the problems and taken positive measures to solve them.

### Acknowledgements

This research was sponsored by the National Natural Science Foundation of China (Grant No. 42171192). The support is gratefully acknowledged. I would like to thank Dr. Linshan Huang, for data processing of Beijing city.